\documentclass[printer]{aa}

\usepackage{amsmath}
\usepackage[T1]{fontenc}
\usepackage[varg]{txfonts}
\usepackage{graphicx}
\usepackage{color}
\usepackage{lineno}
\usepackage{natbib}

\bibpunct{(}{)}{;}{a}{}{,}

\newcommand{\Pm}{\mathrm{Pm}}
\newcommand{\Rey}{\mathrm{Re}}
\newcommand{\Rm}{\mathrm{Rm}}

\newcommand{\tabhe}{\parbox[0pt][2em][c]{0cm}{}}

\begin{document}

\title{Magnetic Field Amplification in Young Galaxies}

\author{J.~Schober \inst{\ref{Heidelberg}}
\and D.~R.~G.~Schleicher \inst{\ref{Goettingen}}
\and R.~S.~Klessen \inst{\ref{Heidelberg}}}

 \institute{Universit\"at Heidelberg, Zentrum f\"ur Astronomie, Institut f\"ur Theoretische Astrophysik, Albert-Ueberle-Strasse\ 2, D-69120 Heidelberg, Germany \label{Heidelberg}
\and Georg-August-Universit\"at G\"ottingen, Institut f\"ur Astrophysik, Friedrich-Hund-Platz 1, D-37077 G\"ottingen, Germany \label{Goettingen}}

\date{\today}

\abstract{The Universe at present is highly magnetized, with fields of the order of a few $10^{-5}$ G and coherence lengths larger than 10 kpc in typical galaxies like the Milky Way.}
{We propose that the magnetic field was amplified to this values already during the formation and the early evolution of galaxies. Turbulence in young galaxies is driven by accretion as well as by supernova (SN) explosions of the first generation of stars. The small-scale dynamo can convert the turbulent kinetic energy into magnetic energy and amplify very weak primordial seed fields on short timescales. Amplification takes place in two phases: in the kinematic phase the magnetic field grows exponentially, with the largest growth rate on the smallest non-resistive scale. In the following non-linear phase the magnetic energy is shifted towards larger scales until the dynamo saturates on the turbulent forcing scale.}
{To describe the amplification of the magnetic field quantitatively we model the microphysics in the interstellar medium (ISM) of young galaxies and determine the growth rate of the small-scale dynamo. We estimate the resulting saturation field strengths and dynamo timescales for two turbulent forcing mechanisms: accretion-driven turbulence and SN-driven turbulence. We compare them to the field strength that is reached, when only stellar magnetic fields are distributed by SN explosions.}
{We find that the small-scale dynamo is much more efficient in magnetizing the ISM of young galaxies. In the case of accretion-driven turbulence a magnetic field strength of the order of $10^{-6}~\mathrm{G}$ is reached after a time of $24-270~\mathrm{Myr}$, while in SN-driven turbulence the dynamo saturates at field strengths of typically $10^{-5}~\mathrm{G}$ after only $4-15~\mathrm{Myr}$. This is considerably shorter than the Hubble time.}
{Our work can help to understand why present-day galaxies are highly magnetized.}


\maketitle


\section{Introduction} 
\label{Sec_Introduction}

The present-day Universe is filled with magnetic fields. Observations show that galaxies \citep{BeckEtAl1999,Beck2011} and stars \citep{DonatiLandstreet2009,Reiners2012} are strongly magnetized and there are also hints towards weak magnetic fields in the intergalactic medium \citep{KimEtAl1989,Kronberg1994,NeronovSemikozBanafsheh2013}. The origin of these strong fields remains an unsolved problem in astrophysics. \\
Local spiral galaxies have typical turbulent field components of $(2-3)\times10^{-5}$ G within the arms and bars, while a field of $(5-10)\times10^{-5}$ G is observed in the central starburst regions. Moreover, these fields appear to be coherent on scales larger than 10 kpc, which is the same order of magnitude as the size of the galaxy. The magnetic energy in the galactic interstellar medium is thus approximately in equipartition with the thermal energy and the energy in cosmic rays. The field in the interarm region is usually ordered and has a strength of the order of $(1-1.5)\times10^{-5}$ G \citep{Beck2011}. Also dwarf irregular galaxies have magnetic fields, however they appear not to be ordered on large scales and have a lower strength of $\leq 4\times10^{-6}$ G \citep{ChyzyEtAl2011}.\\
New observations indicate that even highly redshifted galaxies have magnetic field strengths comparable to present-day galaxies \citep{BernetEtAl2008}. For instance the rotation measure, a quantity depending on the magnetic field along the line of sight, is constant up to redshifts of roughly 5 \citep{HammondRobishawGaensler2012}. An important tool is moreover the FIR-radio correlation, which can be interpreted as a relation between the star formation rate and the synchrotron radiation of cosmic ray electrons in magnetic fields \citep{SargentEtAl2010, BourneEtAl2011} and appears to be valid until $z\approx 2$ \citep{Murphy2009}. We note however, that one expects a breakdown at higher redshift as a result of inverse Compton scattering with cosmic microwave background photons \citep{SchleicherBeck2013}. Observations of the intergalactic medium provide further information on primordial seed fields. Detailed analysis of the CMB temperature bispectrum using data from the PLANCK satellite gives an upper limit of the magnetic field strength of the order of a few nG on the Mpc scale \citep{ShiraishiEtAl2012}. The increasing evidence for magnetic fields in highly redshifted galaxies and the intergalactic medium indicates an early generation of the magnetic fields.\\
Theoretically the first seed fields might already have been generated in the very early Universe during inflation leading to a field strength of $B_0\approx10^{-34}$-$10^{-10}$ G on a scale of $1$ Mpc \citep{TurnerWidrow1988}.
 Another generation mechanism are first order phase transitions. \citet{Sigl1997} predict a field strength of $B_0 \approx 10^{-29} ~\mathrm{G}$ from the electroweak phase transition and $B_0 \approx 10^{-20} ~\mathrm{G}$ from the QCD phase transition on a scale of $10$ Mpc. The correlation length of the primordial magnetic seed fields has been shown to depend crucially on the initial properties of the field, e.g.~on the amount of magnetic helicity \citep{BanerjeeJedamzik2004}. We can determine the typical strength of a statistical seed field by writing the magnetic energy as $B_0^2/(8 \pi)$. Magnetic fields can also be generated as a result of the Biermann term in the generalized Ohm's law, which takes into account the different behavior of the electron and ion fluid \citep{Biermann1950,KulsrudZweibel2008}. A typical field strength resulting from this so-called ``Biermann battery'' is $10^{-19}~\mathrm{G}$ \citep{Xu2008}. Recently, \citet{Schlickeiser2012} has shown that a turbulent magnetic field can be generated in plasma fluctuations within an unmagnetized non-relativistic medium. From this effect we would expect typical seed fields of a few $10^{-10}$ G within the first galaxies. Although this exceeds the resulting field strengths of other generation mechanisms it still cannot explain the typical values in local galaxies. Thus, amplification processes need to take place.\\
A very efficient mechanism to amplify weak seed fields is the small-scale or turbulent dynamo, which converts kinetic energy from turbulence into magnetic energy by randomly stretching and twisting the field lines. The magnetic energy grows exponentially in the kinematic phase{, while the growth rate is largest on the resistive scale} \citep{Kazantsev1968,Subramanian1997, BrandenburgSubramanian2005}. There are two dimensionless parameters, which control the efficiency of this process \citep{SchoberEtAl2012.1,BovinoEtAl2013}: the hydrodynamic Reynolds number
\begin{eqnarray}
  \Rey = \frac{V L}{\nu},
\label{Re}
\end{eqnarray}
where $V$ is the turbulent velocity on the outer scale of the inertial range $L$ and $\nu$ is the viscosity, and the magnetic Reynolds number 
\begin{eqnarray}
  \Rm = \frac{V L}{\eta},
\label{Rm}
\end{eqnarray}
with $\eta$ being the magnetic resistivity. The ratio of the Reynolds numbers defines the magnetic Prandtl number
\begin{eqnarray}
  \Pm \equiv \frac{\Rm}{\Rey} = \frac{\nu}{\eta}.
\label{Pm}
\end{eqnarray}
Furthermore, the dynamo growth rate depends on the type of the turbulence ranging from incompressible Kolmogorov turbulence \citep{Kolmogorov1941} to highly compressible Burgers turbulence \citep{Burgers1948}. Eventually the magnetic field is strong enough for back reactions to occur on the gas and the non-linear growth sets in \citep{SchekochihinEtAl2002.2}. In this phase the magnetic energy is transported towards larger scales. Now the evolution no longer depends on the Reynolds and Prandtl numbers, but still on the type of turbulence \citep{SchleicherEtAl2013}. The non-linear phase comes to an end, when the magnetic field reaches saturation on the turbulent forcing scale $L$.\\
Turbulence is driven efficiently for the first time in the history of the Universe when dark matter halos become massive enough that gas begins to cool efficiently and flows into the potential wells of dark matter halos. This leads to the formation of the first generation of stars and the subsequent build-up of galaxies. The first stars form at redhifts between 20 and 15 within primordial minihalos, which have typical masses of more than $10^5$ solar masses ($M_\odot$) (see e.g. \citet{AbelEtAl2002,BrommLarson2004,ClarkEtAl2011}). In recent publications it was shown, numerically as well as semi-analytically, that during the formation of the first stars, dynamically important magnetic fields can be generated by the small-scale dynamo on short timescales \citep{SurEtAl2012,TurkEtAl2012,SchoberEtAl2012.2}. According to the theory of hierarchical structure formation the first galaxies, also called protogalaxies, form at redshifts smaller than about 10 in massive dark matter halos with more than $10^7~M_\odot$ \citep{GreifEtAl2008,BrommEtAl2009}. In young galaxies accretion as well as the penetration of supernovae (SN) shocks through the gas generate turbulence, which initiates small-scale dynamo action \citep{BeckEtAl2012, LatifEtAl2013}. \\
In this paper, we follow the evolution of the magnetic field in an initially weakly magnetized young galaxy. Because the true dynamical nature of the first galaxies is not well known, we adopt two simplified complementary models: a spherical galaxy as well as a disk-like system, both with constant density and temperature. We model microphysical processes, such as the diffusion of the kinematic and magnetic energy, in order to find the magnetohydrodynamical (MHD) quantities, which determine the growth rate of the small-scale dynamo. Turbulence can be generated by accretion flows into the center of the halo, for which we estimate the typical Reynolds numbers. Then we follow the evolution of the magnetic field strength in the kinematic and the non-linear phase, until saturation on the driving scale of the turbulence is reached. Also stellar feedback, in particular SN explosions, influences the evolution of the magnetic field. On the one hand supernovae distribute stellar magnetic fields in the interstellar medium (ISM) \citep{Rees1987}, on the other hand they drive turbulence, which again leads to dynamo action \citep{BalsaraEtAl2004}. We compare the resulting magnetic field strengths from both mechanisms with the field strength gained by an accretion-driven small-scale dynamo.\\
The outline of the paper is as follows: In section \ref{Sec_Model} we describe our model. We determine the values of viscosity and magnetic diffusivity in the interstellar medium and estimate the evolution of SN explosions. Driving mechanisms of turbulence are discussed in general. In the last part of this section we summarize the main points of a mathematical description of magnetic field amplification by the small-scale dynamo. The kinematic phase described by the so-called Kazantsev theory and a model for the non-linear growth phase are introduced. In section \ref{Sec_Results} we present our results for the evolution of the magnetic field in the different types of models. First we discuss the generation of turbulence by accretion and the resulting efficiencies of the dynamo, i.e.~the saturation magnetic field strength and the time until saturation occurs. Second, we analyze the effect of stellar feedback. We compare the efficiency of distributing stellar magnetic fields by SN with the one of the SN-driven turbulent dynamo. We draw our conclusions in section \ref{Sec_Conclusions}.

\section{Modeling Physical Processes in a Protogalaxy}
\label{Sec_Model}

\subsection{General Aspects}
\label{Sec_Model_General}

The nature of young galaxies is still an active topic of research (see \citet{BrommYoshida2011} for a review). For our order of magnitude estimate of the magnetic field evolution we use a very simplified model, with the choice of parameters being motivated from numerical simulations \citep{GreifEtAl2008,BrommEtAl2009,LatifEtAl2013}. We are interested in massive protogalactic objects at redshifts of roughly 10. \\
In our model we assume a mean particle density of
\begin{eqnarray}
  n = 10~\mathrm{cm}^{-3},
\end{eqnarray}
and a temperature of
\begin{eqnarray}
  T = 5\times 10^{3}~\mathrm{K}.
\end{eqnarray}
The density as well as the temperature are, as first approximation, constant throughout the whole galaxy. For simplicity we take into account a gas that only consists of hydrogen, which is at the given values of $n$ and $T$ mostly ionized.\\
The mean shape of the primordial galaxies differs most probably from the one of present-day galaxies. Due to a significant amount of angular momentum the protogalaxies may form in a spherical way and develop a more disk-like structure at later stages. To account for the unknown typical shape, we model two extreme cases, a spherical and a disk-like galaxy, which have the same gas mass.

\paragraph{Spherical Galaxy}
In the case of a spherical protogalaxy we assume the radius to be 
\begin{eqnarray}
  R_\mathrm{sph} = 10^{3}~\mathrm{pc}.
\label{Rsph}
\end{eqnarray}
As within this radius the density as well as the temperature are constant we find a total mass of the baryonic gas of
\begin{eqnarray}
  M \approx 10^{9}~M_\odot.
\end{eqnarray}

\paragraph{Disk-Like Galaxy}
As our second fiducial model we use a galaxy with disk scale height of ten percent of the radius, i.e.
\begin{eqnarray}
  H_\mathrm{disk} = 0.1~R_\mathrm{disk}.
\end{eqnarray}
With the condition that the gas mass of the disk needs to be the same as in the spherical case, the disk radius is
\begin{eqnarray}
  R_\mathrm{disk} \approx 2.4\times10^{3}~\mathrm{pc}.
\label{Rdisk}
\end{eqnarray}

\subsection{Microphysics in the ISM}
\label{microphysics}

\begin{figure}[hbtp]
  \centering
  \includegraphics[width=0.48\textwidth]{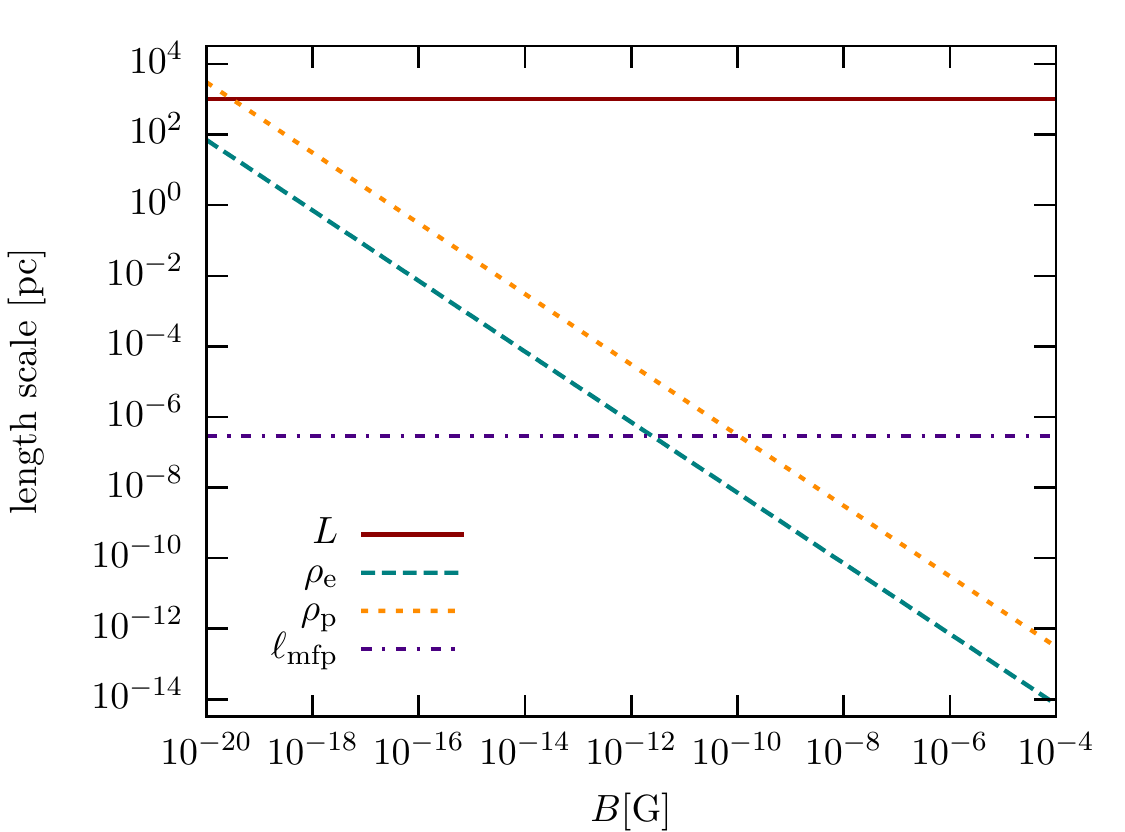}
\caption{The gyro-radii of electrons and ions $\rho_\mathrm{e}$ and $\rho_\mathrm{p}$ as a function of magnetic field strength compared to the typical macroscopic scale $L\approx10^3$ pc and the mean-free path $\ell_\mathrm{mfp}$. Within our fiducial case for the density and the temperature the electron fluid becomes magnetized at a magnetic field strength of roughly $10^{-12}$ G, the ion fluid at $10^{-10}$ G.}
\label{LengthScales}
\end{figure}

As the temperature in the primordial ISM is very high, we can assume the gas to be (at least partially) ionized. We thus need to deal with the full plasma equations, i.e.~the continuity, the momentum and the energy equations for both the ions and the electrons. Closures of these equations were found by \citet{Braginskii1965}, who used the Chapman-Enskog scheme \citep{ChapmanCowling1953}. The closure is based on the assumption that the macroscopic scale of the plasma is large compared to the mean-free path 
\begin{eqnarray}
  \ell_\mathrm{mfp} = \frac{1}{n r_\mathrm{c}^2},
\label{mfp}
\end{eqnarray}
or compared to the gyro-radii of the electrons and the ions
\begin{eqnarray}
  \rho(B) = \frac{(2 m_\mathrm{s} k T)^{1/2} c}{e B}.
\label{gyro}
\end{eqnarray}
Here $r_\mathrm{c}=e^2/(k T)$ is the distance of closest particle approach with $e$ being the elementary charge and $k$ the Boltzmann constant. The mass of the species is labeled $m_\mathrm{s}$, where the s stands for electrons (e) or protons (p), and $c$ is the speed of light. Further, we use here the thermal velocity $(2 k T / m_\mathrm{s})^{1/2}$ and assume that the temperatures of the ions and electrons are equal ($T_\mathrm{e}=T_\mathrm{p} \equiv T$).
In principle, the components of a plasma can have unequal temperatures as during plasma heating the different fluids are heated differently. However, after a certain time $t_\mathrm{eq}$, an equilibrium will be reached. The electron-proton equilibrium time can be computed by \citep{Spitzer1956}
\begin{eqnarray}
  t_\mathrm{eq} = \frac{3 m_\mathrm{e} m_\mathrm{p} k^{3/2}}{8 (2 \pi)^{1/2} n Z_\mathrm{e}^2 Z_\mathrm{p}^2 e^4 \mathrm{ln}(\Lambda)} \left(\frac{T_\mathrm{e,0}}{m_\mathrm{e}} + \frac{T_\mathrm{p,0}}{m_\mathrm{p}}\right)^{3/2},
\end{eqnarray}
where $Z_\mathrm{s}$ is the charge of species s, $T_\mathrm{s,0}$ its initial temperature and the Coulomb logarithm is defined by
\begin{eqnarray}
  \mathrm{ln}(\Lambda) &\approx & 6.6 - 0.5~\mathrm{ln}\left(\frac{n}{10^{14}~\mathrm{cm}^{-3}}\right)    \nonumber \\  
                       &        & + 1.5~\mathrm{ln}\left(\frac{k T}{1.6\times 10^{-12}~\mathrm{erg}}\right).
\end{eqnarray}
If we assume $T_\mathrm{e,0}$ and $T_\mathrm{p,0}$ to be extremely different, e.g. $T_\mathrm{e,0} = 10^3~T_\mathrm{p,0}$, the typical $t_\mathrm{eq}$ for our model is of the order of 440 yr. It will be shown later that this is way below the typical dynamo timescales, which can be up to many Myr. Thus, the electron and the proton temperature can be assumed to be equal in our calculation. \\
A comparison of the length scales (\ref{mfp}) and (\ref{gyro}) in our model can be found in figure \ref{LengthScales}. When the gyro-radius becomes smaller than the mean-free path, the magnetic field dominates the dynamics of the plasma, i.e.~it becomes ``magnetized''. In our model the electron fluid becomes magnetized at a magnetic field strength of roughly $10^{-12}$ G, the ion fluid at $10^{-10}$ G.  

\subsubsection{Viscosity}
\begin{figure}[hbtp]
  \centering
  \includegraphics[width=0.48\textwidth]{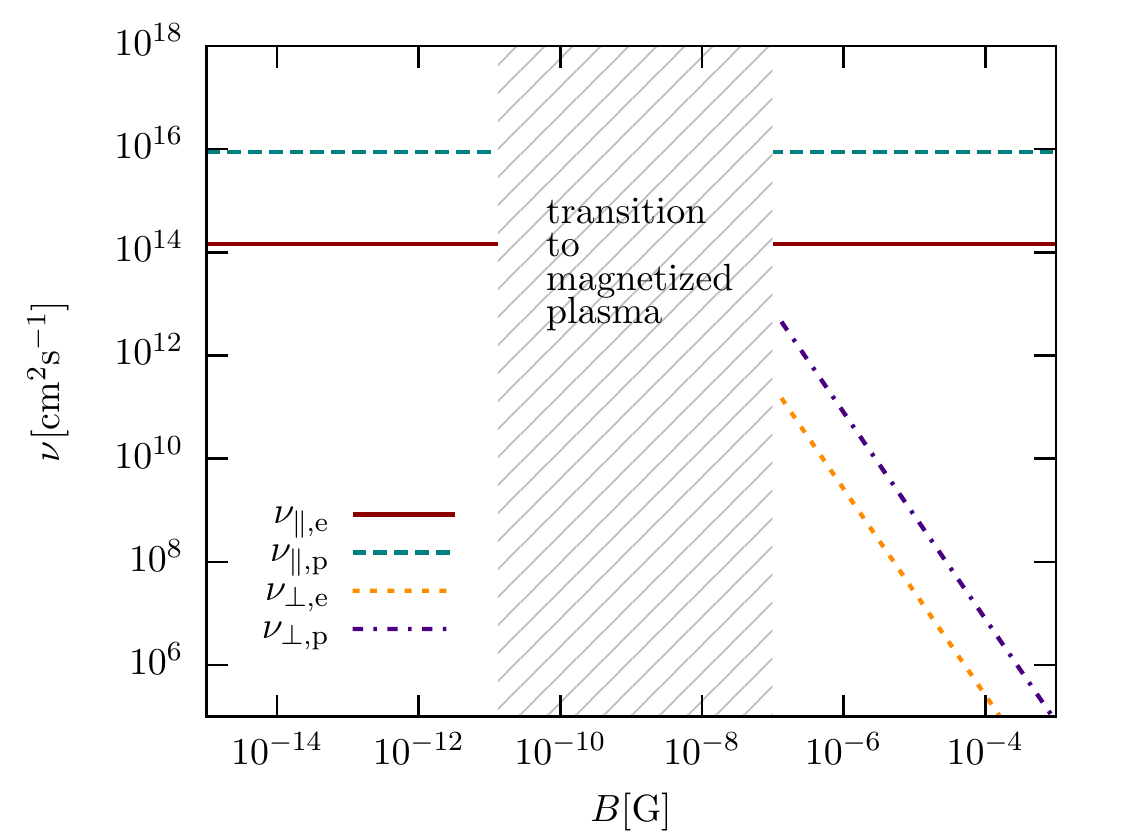}
\caption{The kinematic viscosity parallel ($\nu_\parallel$) and perpendicular to the magnetic field lines ($\nu_\perp$) as a function of magnetic field strength $B$. We show the results for the electron as well as for the ion fluid. The range between $10^{-11}$ G and $10^{-7}$ G is not shown, as here the transition from the unmagnetized to a magnetized plasma takes place.}
\label{Viscosity}
\end{figure}

In the transition from an unmagnetized to a magnetized state, the plasma becomes anisotropic, i.e.~certain physical quantities then depend on their relative orientation to the magnetic field direction. \\
In the unmagnetized case the kinematic viscosities for electrons and ions obtained from the Chapman-Enskog closure scheme are \citep{Braginskii1965}
\begin{eqnarray}
  \nu_{\parallel,\mathrm{e}} & = & 0.73 \frac{\tau_\mathrm{e} k T}{m_\mathrm{e}} = 1.4\times10^{14}~\mathrm{cm}^2\mathrm{s}^{-1} 
\label{nu_parallel_e}
\end{eqnarray}
\begin{eqnarray}
  \nu_{\parallel,\mathrm{p}} & = & 0.96 \frac{\tau_\mathrm{p} k T}{m_\mathrm{p}} = 8.7\times10^{15}~\mathrm{cm}^2\mathrm{s}^{-1}
\label{nu_parallel_p}
\end{eqnarray}
with the collision times for electrons and ions 
\begin{eqnarray}
  \tau_{\mathrm{e}} & = & \frac{6 \sqrt{2} \pi^{3/2} \sqrt{m_\mathrm{e}} (k T)^{3/2}}{16 \pi^2 \mathrm{ln}(\Lambda) e^4 n}\\
  \tau_{\mathrm{p}} & = & \frac{12 \pi^{3/2} \sqrt{m_\mathrm{p}} (k T)^{3/2}}{16 \pi^2 \mathrm{ln}(\Lambda) e^4 n}.
\end{eqnarray}
In the presence of a strong magnetic field the viscosity becomes anisotropic and one has to distinguish between the viscosity along (parallel to) and the one perpendicular to the magnetic field lines. While the parallel viscosity stays the same as in the unmagnetized case (e.g.~equations \ref{nu_parallel_e} and \ref{nu_parallel_p}), the viscosity perpendicular to the field is given by \citep{Simon1955}
\begin{eqnarray}
  \nu_{\perp,\mathrm{e}}(B) & = & 0.51 \frac{k T}{\Omega_\mathrm{e}(B)^2 \tau_\mathrm{e} m_\mathrm{e}}\\
  \nu_{\perp,\mathrm{p}}(B) & = & \frac{3 k T}{10 \Omega_\mathrm{p}(B)^2 \tau_\mathrm{p} m_\mathrm{p}}
\end{eqnarray}
with the gyro-frequencies 
\begin{eqnarray}
  \Omega_{\mathrm{e}}(B) & = & \frac{e B}{m_\mathrm{e} c}\\
  \Omega_{\mathrm{p}}(B) & = & \frac{e B}{m_\mathrm{p} c}.
\end{eqnarray}
Diffusion perpendicular to the magnetic field lines is also known as ``Bohm diffusion''.\\
The different viscosities as a function of density are shown in figure \ref{Viscosity}. Note that the perpendicular viscosity becomes only valid when the plasma is magnetized, i.e.~when the gyro-radius becomes smaller than the mean-free path. According to figure \ref{LengthScales} this is the case above a magnetic field strength of $10^{-11}$ G for the electrons and $10^{-9}$ G for the ions. Thus, the most important part of the viscosity is the parallel one and we will ignore the perpendicular part, which decreases proportional the $1/B^2$, from now on. Furthermore, the viscosity of the ions exceeds the electron viscosity by roughly two orders of magnitude. This is caused by the fact that the ions carry the largest part of the momentum. In total, the parallel viscosity of the ions is the crucial quantity and we will refer from now on to 
\begin{eqnarray}
  \nu \equiv \nu_{\parallel,\mathrm{p}} \approx 8.7\times 10^{15}~\mathrm{cm}^2\mathrm{s}^{-1}.
\end{eqnarray}

\subsubsection{Magnetic Diffusivity}

For the parallel conductivity the closure scheme yields \citep{Spitzer1956}
\begin{eqnarray}
  \sigma_{\parallel} = 1.96 \frac{n e^2 \tau_\mathrm{e}}{10^{-7} c^2 m_\mathrm{e}}
\end{eqnarray}
and for the conductivity perpendicular to the magnetic field
\begin{eqnarray}
  \sigma_{\perp} = 0.51 \sigma_{\parallel}.
\end{eqnarray}
The conductivity perpendicular to the magnetic field lines is, contrary to the case of viscosity, no function of the magnetic field strength. The difference between the parallel and the perpendicular component of the conductivity is just approximately a factor of two. Usually, $\sigma_{\parallel}$ is used to determine the magnetic diffusivity $\eta$ of a plasma. We thus find 
\begin{eqnarray}
  \eta = \frac{1}{\sigma_{\parallel}} = 37.8~\mathrm{cm}^2\mathrm{s}^{-1},
\label{eta}
\end{eqnarray}
which is also known as ``Spitzer resistivity''. 

\subsubsection{Magnetic Prandtl Number}
With these values of viscosity and resistivity the magnetic Prandtl number (see equation \ref{Pm}) is
\begin{eqnarray}
  \Pm \approx 3.7\times10^{14}.
\end{eqnarray}

\subsection{Turbulence}
\label{SubSec_Turbulence}

\subsubsection{Generation of Turbulent Motions by Accretion}

Structure formation is always associated with accretion. In order to build up the first stars and galaxies, gas flows into the potential wells of dark matter halos, where it gets compressed and cools. The potential energy released during that process in parts gets converted into turbulent kinetic energy \citep{KlessenHennebelle2010}. Simulations by \citet{GreifEtAl2008} of atomic cooling halos show that two types of accretion occur: in the so-called ``hot accretion'' mode gas is accreted directly from the intergalactic medium, while in the ``cold accretion'' mode gas is cooled down and flows into the central regions of the halo at high velocities \citep{DekelEtAl2009,NelsonEtAl2013}.

\subsubsection{Generation of Turbulent Motions by Supernova Explosions}

Once stars have formed, their feedback strongly influences the ISM in galaxies in terms of ionizing radiation and at later stages by SN explosions, which are especially important for the generation of turbulence. \\
In order to calculate the corresponding energy input, we need to estimate the rate of SN explosions. The star formation rate ($SFR$) is proportional to the mass density $\rho = n m$ over the free-fall time $t_\mathrm{ff} = (3 \pi/(32 G \rho))^{1/2}$ \citep{MacLowKlessen2004,McKeeOstriker2007}:
\begin{eqnarray}
  SFR \propto \frac{\rho}{t_\mathrm{ff}}.
\label{SFR}
\end{eqnarray}
From the $SFR$ we can estimate the supernova rate ($SNR$). For this we divide the $SFR$ by the typical mass of a star that results in a SN ($10~M_\odot$). As not all the gas goes into stars and not all the stars are massive enough to end in a SN we introduce an efficiency factor $\alpha$:
\begin{eqnarray}
  SNR \approx  \alpha \frac{\rho}{t_\mathrm{ff}\cdot10~M_\odot}.
\label{SNR}
\end{eqnarray}
The number of supernovae within the whole galaxy with a volume $V_\mathrm{gal}$ and a time interval $t$ is then given by
\begin{eqnarray}
  N_\mathrm{SN}(t) = SNR~V_\mathrm{gal}~t,
\end{eqnarray}
where we assume that the $SNR$ stays constant over time. In general the SN rate is expected to change with time, however modeling this time dependency goes beyond the scope of this work.

\subsection{Turbulent Magnetic Field Amplification}

\subsubsection{Kinematic Small-Scale Dynamo}
The induction equation,
\begin{eqnarray}
  \frac{\partial \textbf{B}}{\partial t} = \nabla\times\left(\textbf{v}\times\textbf{B} - \eta\nabla\times\textbf{B}\right),
\label{induction}
\end{eqnarray} 
describes the time evolution of a magnetic field $\textbf{B}$, where $\textbf{v}$ is the velocity and $\eta$ the magnetic diffusivity (\ref{eta}). An arbitrary magnetic field can, in general, be separated into a mean component $\textbf{B}_0$ and a fluctuating component $\delta \textbf{B}$ with 
\begin{eqnarray}
  \textbf{B} = \textbf{B}_0 + \delta \textbf{B}.
\label{B0deltaB}
\end{eqnarray} 
Substituting (\ref{B0deltaB}) into the induction equation leads to two equations: an equation for the large-scale field evolution and the Kazantsev equation \citep{Kazantsev1968,BrandenburgSubramanian2005}, which describes the small-scale evolution of the field. \\
The derivation of the Kazantsev equation is based on the assumption that the fluctuations of the magnetic field as well as the fluctuations of the velocity field are homogeneous and isotropic even if the mean fields are not isotropic. Furthermore, the fluctuations are assumed to be Gaussian with a zero mean and the velocity fluctuations are thought to be $\delta$-correlated in time. For simplicity, any helicity of the magnetic field is neglected. With these assumptions the Kazantsev equation is \citep{Kazantsev1968}
\begin{eqnarray}
  -\kappa_\mathrm{diff}(r)\frac{\mathrm{d}^2\psi(r)}{\mathrm{d}^2r} + U(r)\psi(r) = -\Gamma \psi(r).
\label{Kazantsev}
\end{eqnarray}
The eigenfunctions of this equation are related to the longitudinal correlation function of the magnetic fluctuations $M_L(r,t)$ by $M_\mathrm{L}(r,t) \equiv 1/(r^2\sqrt{\kappa_\mathrm{diff}})~\psi(r)~\mathrm{e}^{2\Gamma t}$. We call $\Gamma$ the growth rate of the small-scale magnetic field. The function $\kappa_\mathrm{diff}$ is the magnetic diffusion coefficient, which contains besides the magnetic diffusivity $\eta$ also a scale-dependent turbulent diffusivity. $U$ is called the ``potential'' of the Kazantsev equation. Both $\kappa_\mathrm{diff}$ and $U$ depend only on the correlation function of the turbulent velocity field and the magnetic diffusivity \citep{Subramanian1997,SchoberEtAl2012.1}. \\
The correlation function of the turbulent velocity field in turn depends on the different types of turbulence, which can be distinguished by the slope of the velocity spectrum $\vartheta$ in the inertial range, where
\begin{eqnarray}
  \delta v \propto \ell^{\vartheta}.
\end{eqnarray}
Here $\delta v$ is the velocity of the fluctuations on the scale $\ell$. The range of $\vartheta$ goes from incompressible Kolmogorov turbulence with $\vartheta = 1/3$ \citep{Kolmogorov1941} to highly compressible Burgers turbulence with $\vartheta = 1/2$ \citep{Burgers1948}. The gas motions during structure formation have high Mach numbers and thus the gas gets strongly compressed within shocks. Observations within present-day molecular clouds by \citet{Larson1981} show that the slope of the turbulence spectrum is $\vartheta \approx 0.38$ and thus deviates from Kolmogorov turbulence. However, other studies \citep{SolomonEtAl1987,OssenkopfMacLow2002,HeyerBrunt2004} find a slope of roughly 0.5, whereas \citet{RomanDuvalEtAl2011} show that the variance of $\vartheta$ is very large. For our fiducial model we choose a value of $\vartheta = 0.4$, which lies in between the extremes. \\
With a model for the turbulent correlation function, the Kazantsev equation (\ref{Kazantsev}) can be solved with the WKB-approximation for very large and low Pm. This method is named after Wentzel, Kramers and Brillouin and is used to find approximative solutions for Schr\"odinger-type differential equations. In our model we are in the limit of the very high Pm, where \citet{SchoberEtAl2012.1} find the growth rate
\begin{eqnarray}
  \Gamma = \frac{(163-304\vartheta)}{60}\frac{V}{L}Re^{(1-\vartheta)/(1+\vartheta)}.
\label{Gamma}
\end{eqnarray}
Here $V$ is the typical velocity on the largest scale of the turbulent eddies of size $L$. By solving the Kazantsev equation (\ref{Kazantsev}) numerically, \citet{BovinoEtAl2013} have recently confirmed that equation (\ref{Gamma}) describes the growth rate of the dynamo in the limit large Pm. For our fiducial model with $\vartheta = 0.4$ the growth rate thus scales with $\Rey^{0.43}$.

\subsubsection{Non-Linear Small-Scale Dynamo}

As soon as the magnetic energy is comparable to the kinetic energy of the turbulence on the viscous scale the exponential growth comes to an end. We label this point in time $t_\nu$. The dynamo is then saturated on the viscous scale and the non-linear growth begins. In this phase the magnetic energy $E_\mathrm{mag}$ on the scale of fastest amplification $\ell_\mathrm{a}$ evolves as \citep{SchekochihinEtAl2002.2}
\begin{eqnarray}
  \frac{\mathrm{d}}{\mathrm{d} t} E_\mathrm{mag}(t) = \Gamma_\mathrm{nl}(t) E_\mathrm{mag}(t) - 2 \eta k_\mathrm{rms}^2 E_\mathrm{mag}(t)
\label{NLevolution}
\end{eqnarray}
with the non-linear growth rate
\begin{eqnarray}
  \Gamma_\mathrm{nl}(t) \approx \frac{v_{\ell_\mathrm{a}}(t)}{\ell_\mathrm{a}(t)},
\end{eqnarray}
and
\begin{eqnarray}
  k_\mathrm{rms}^2(t) = \frac{1}{E_\mathrm{mag}} \int_0^\infty \mathrm{d}k~k^2 \left(\frac{1}{2}\int \mathrm{d} \Omega_\textbf{k} \langle|\textbf{B}(t,\textbf{k})|^2\rangle\right).
\end{eqnarray}
\citet{SchleicherEtAl2013} find that further evaluation of equation (\ref{NLevolution}) yields
\begin{eqnarray}
  \frac{\mathrm{d}}{\mathrm{d} t} E_\mathrm{mag} \propto E_\mathrm{mag}^{1+(\vartheta-1)/(2\vartheta)}.
\end{eqnarray}
Thus, in the case of Kolmogorov turbulence with $\vartheta=1/3$ the magnetic energy grows linear in time, while it grows quadratically in case of Burgers turbulence with $\vartheta=1/2$. In our fiducial model, where we assume $\vartheta=0.4$, we find $E_\mathrm{mag} \propto t^{4/3}$ on $\ell_\mathrm{a}$. \\ 
In the non-linear phase the dynamo process shifts the magnetic energy to larger scales with the peak scale evolving as
\begin{eqnarray}
  \ell_{\mathrm{p}}(t) = \ell_\nu + \left(\frac{V}{L^\vartheta}\left(t-t_\nu\right)\right)^{1/(1-\vartheta)}.
\label{lp}
\end{eqnarray}
From the peak scale to larger scales we assume the spectrum to drop off with the Kazantsev slope. By this we can determine the magnetic field on the forcing scale $L$ at each point in time as
\begin{eqnarray}
  B_{\mathrm{L}}(t) = B_{\ell_\mathrm{p}}(t)~\left(\frac{\ell_\mathrm{p}(t)}{L}\right)^{5/4}.
\end{eqnarray}
The non-linear growth phase comes to an end, when saturation on the turbulent forcing scale is achieved. Now the spectrum of the magnetic energy density scales as the one of the kinetic energy density.

\subsubsection{Saturation Magnetic Field Strength from Dynamo Amplification}

\begin{figure}[hbtp]
  \centering
  \includegraphics[width=0.48\textwidth]{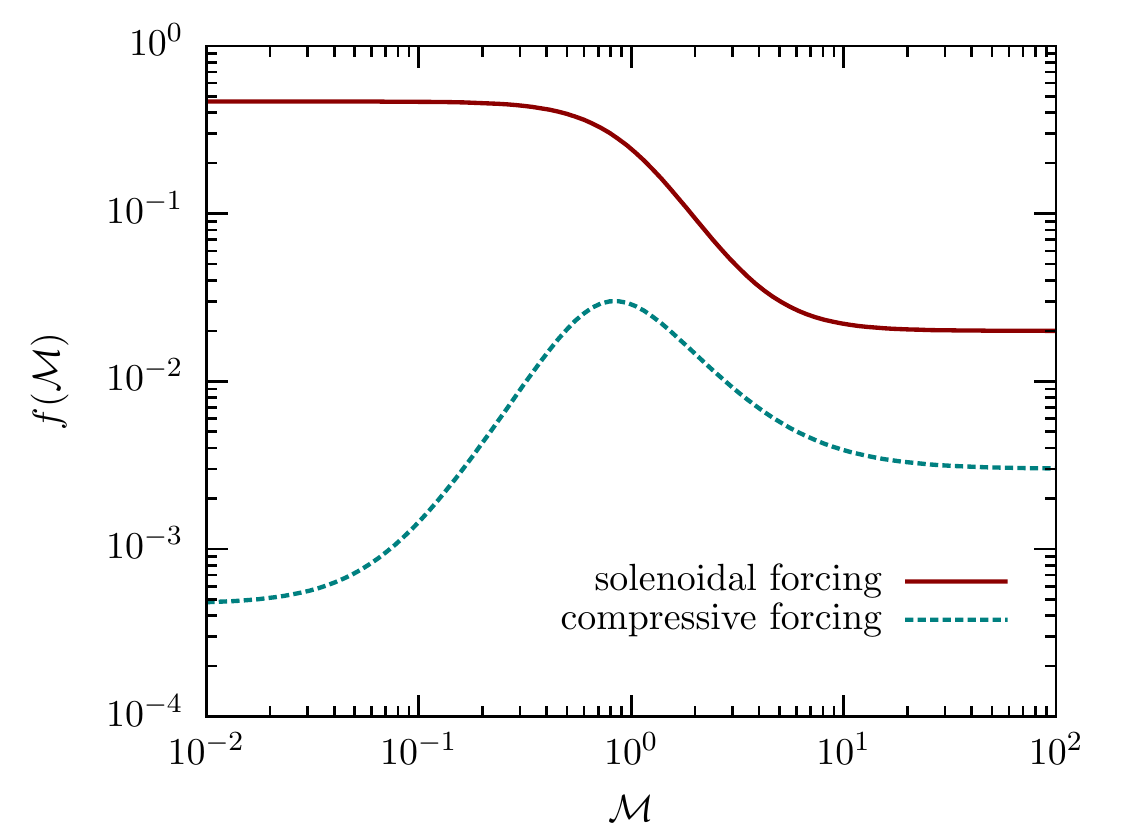}
\caption{The ratio of magnetic over turbulent kinetic energy at saturation $f(\mathcal{M})$ as a function of the Mach number $\mathcal{M}$. We present fits for solenoidal (solid line) and compressive forcing (dashed line) of the turbulence from the driven MHD simulations by \citet{FederrathEtAl2011.2}.}
\label{SaturationEfficiency}
\end{figure}

A turbulent dynamo can amplify magnetic fields at most to equipartition with the turbulent kinetic energy. However, high-resolution simulations by \citet{FederrathEtAl2011.2} show that only a certain fraction $f$ of the turbulent kinetic energy can be transformed into magnetic energy. This fraction depends on the type of forcing as well as on the Mach number $\mathcal{M}$. We show $f(\mathcal{M})$ for solenoidal and compressive forcing of turbulence in figure \ref{SaturationEfficiency}. Note, that the efficiency of the small-scale dynamo in case of compressive forcing peaks at a Mach number of 1, i.e.~at the transition from the subsonic to the supersonic regime. At this point shocks appear, which generate solenoidal motions that are more efficient for dynamo amplification. At larger Mach numbers the efficiency decreases again and appears to become constant. \\
According to \citet{Federrath2010} solenoidal forcing leads to a slope of the turbulence spectrum of $0.43$, while compressive forcing results in $\vartheta \approx 0.47$. For our fiducial model we choose the saturation efficiency of solenoidal driven turbulence, as we assume a spectrum with $\vartheta = 0.4$.
\\
The resulting saturation magnetic field strength on the forcing scale is
\begin{eqnarray}
  B_{L,\mathrm{sat}} = \left(4 \pi \rho\right)^{1/2}~V~f(\mathcal{M})^{1/2},
\end{eqnarray}
where $V$ is again the velocity at the forcing scale. If we scale down the turbulent velocity to the viscous scale by
\begin{eqnarray}
  v_{\nu} = \left(\frac{\ell_{\nu}}{L}\right)^{\vartheta}~V
\end{eqnarray}  
the saturation magnetic field strength on the viscous scale is
\begin{eqnarray}
  B_{\nu,\mathrm{sat}} = \left(4 \pi \rho\right)^{1/2}~\left(\frac{\ell_{\nu}}{L}\right)^{\vartheta}~V~f(\mathcal{M})^{1/2}.
\end{eqnarray}

\subsubsection{Evolution of a Magnetic Field Amplified by the Small-Scale Dynamo}

\begin{figure}[hbtp]
  \centering
  \includegraphics[width=0.48\textwidth]{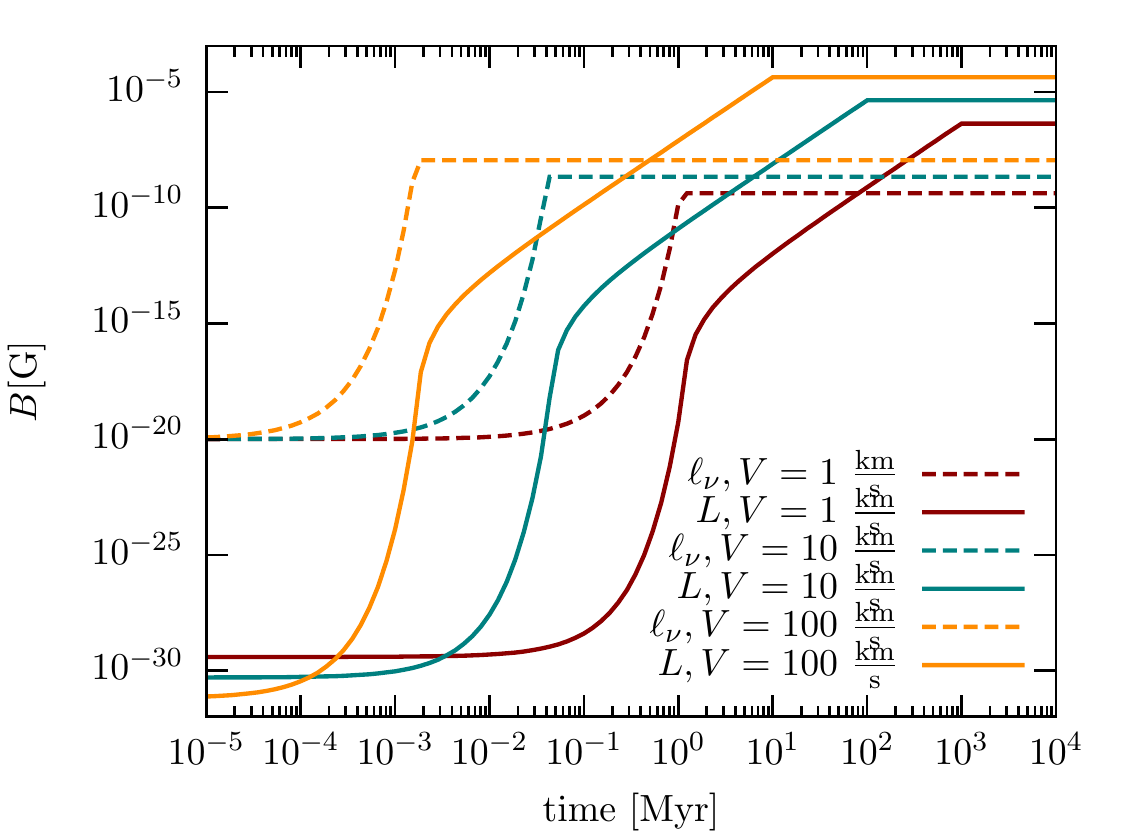}
\caption{The evolution of the magnetic field amplified by the small-scale dynamo. The dashed lines show the evolution on the viscous scale $\ell_\nu$, the solid lines the one on the forcing scale of the turbulence $L$. We use in this plot $L = 10^3$ pc. The different colors indicate different turbulent velocities: the red lines have velocities of $1~\mathrm{km~s}^{-1}$, the green lines $10~\mathrm{km~s}^{-1}$ and the orange lines $100~\mathrm{km~s}^{-1}$. The microphysical quantities are determined in section \ref{microphysics}. The viscosity is $\nu = 8.7\times 10^{15}~\mathrm{cm}^2\mathrm{s}^{-1}$, the magnetic resistivity $\eta = 37.8~\mathrm{cm}^2\mathrm{s}^{-1}$, the density is $n = 10~\mathrm{cm}^{-3}$ and the mean particle mass is $m = 1.6\times 10^{-24}~\mathrm{g}$. The initial magnetic field strength on the viscous scale in this plot is $B_0 = 10^{-20}$ G and the slope of the turbulence spectrum $\vartheta$ is $0.4$.}
\label{Bfield_generic}
\end{figure}

Summarizing the results of this section gives for the magnetic field evolution on the viscous scale
\begin{eqnarray}
  B_{\nu}(t) = \begin{cases} 
                  B_{\nu,0}~\mathrm{exp}(\Gamma t) &  t < t_\nu \\
                  B_{\nu,\mathrm{sat}}             &  t \geq t_\nu,
               \end{cases}
\label{Bnu}
\end{eqnarray}
i.e.~it grows exponentially with rate (\ref{Gamma}) until saturation on the viscous scale is reached at the time $t_\nu$. \\
The field on the turbulent forcing scale evolves as
\begin{eqnarray}
  B_{L}(t) = \begin{cases} 
                B_{\ell_\nu,0}~\mathrm{exp}(\Gamma t)~\left(\frac{\ell_\nu}{L}\right)^{5/4}  &  t < t_\nu \\
    	        (4\pi\rho)^{1/2} V \left(\frac{\ell_\mathrm{p}(t)}{L}\right)^{\vartheta+5/4}f(\mathcal{M})^{1/2}  &  t_\nu \leq t < t_L \\
                B_{L,\mathrm{sat}}  &    t \geq t_L.
             \end{cases}
\label{BL}
\end{eqnarray}
Until the time $t_\nu$ the field grows exponentially in the kinematic phase. For $t \geq t_\nu$ the dynamo is in the non-linear phase, in which the peak of the magnetic spectrum, which is given by equation (\ref{lp}), is shifted towards larger scales. The dynamo is saturated on all scales of the turbulent inertial range including the driving scale for times $t \geq t_L$. \\
The dynamo amplification of a weak magnetic seed field of $10^{-20}$ G is shown in figure \ref{Bfield_generic}. We choose here a forcing scale of $10^3$ pc, which is the radius of the spherical halo considered here, and three different turbulent velocities: $1~\mathrm{km~s}^{-1}$, $10~\mathrm{km~s}^{-1}$ and $100~\mathrm{km~s}^{-1}$. The microphysical quantities are taken from the calculations in the previous sections. In the figure the dashed lines represent the magnetic field strength on the viscous scale, the solid lines the one on the forcing scale.

\section{Magnetic Field Evolution in a Protogalaxy}
\label{Sec_Results}

\subsection{Magnetic Fields from an Accretion-Driven Small-scale Dynamo}

\subsubsection{Forcing Turbulence by Accretion}

\paragraph{Accretion in a Spherical Galaxy}
During the formation of the primordial halo turbulence is generated by accretion \citep{BirnboimDekel2003,SemelinCombes2005,WiseTurkAbel2008,VogelsbergerEtAl2013}. Simulations show that accretion flows have high Mach numbers with respect to the cold gas even in the central regions of the halo \citep{GreifEtAl2008}. The characteristic forcing scale in case of a spherical halo is the radius, where the accretion flow comes to a halt: 
\begin{eqnarray}
  L_\mathrm{acc} \approx R_\mathrm{sph}.
\end{eqnarray}
\citet{LatifEtAl2013} show in their simulation of a nearly isothermal protogalaxy that the Mach number in such environment is roughly 2. Thus, the typical turbulent velocities from accretion are of the order of
\begin{eqnarray}
  V_\mathrm{acc} \approx 2~c_\mathrm{s}~\beta^{1/2},
\label{Vacc_sphere}
\end{eqnarray}
where $c_\mathrm{s} = (\gamma k T/m)^{1/2}$ is the sound speed and we use an adiabatic index $\gamma$ of 5/3. Further, we assume here that only a certain fraction $\beta$ of the kinetic energy of the accretion flows goes into turbulence, with $\beta$ typically depending on the density contrast between the accretion flows and the halo \citep{KlessenHennebelle2010}. Simulations \citep{LatifEtAl2013} indicated that about five percent of the kinetic energy are in turbulent motions, i.e.~$\beta\approx0.05$.\\
The resulting turbulent length scales, velocities and Reynolds numbers for a spherical galaxy are given in table \ref{Table}.

\paragraph{Accretion in a Disk-Like Galaxy}
In the case of a disk-like galaxy we adopt the typical forcing scale of the turbulence by accretion flows to be the scale height
\begin{eqnarray}
  L_\mathrm{acc} \approx H_\mathrm{disk}.
\end{eqnarray}
We further estimate the typical velocity for accretion flows to be of the order of the Kepler velocity in a disk
\begin{eqnarray}
  V_\mathrm{Kepler} \approx \left(G~n ~m ~\pi~R_\mathrm{disk}~ H_\mathrm{disk}\right)^{1/2}.
\label{VKepler}
\end{eqnarray}
If a percentage $\beta$ of the kinetic energy goes into turbulence, the resulting turbulent velocity is given by
\begin{eqnarray}
  V_\mathrm{acc} \approx V_\mathrm{Kepler}~\beta^{1/2}.
\label{Vacc_disk}
\end{eqnarray}
Typical values of the length scale, the velocity scales and the Reynolds numbers with a value of $\beta=0.05$ are summarized in table \ref{Table}.

\begin{table*} 
\centering
     \begin{tabular}{lllll}
      \multicolumn{1}{c}{} & \multicolumn{2}{c}{Accretion-Driven Dynamo} & \multicolumn{2}{c}{SN-Driven Dynamo}  \\
      \cline{2-3} \cline{4-5}
      \parbox[0pt][2.5em][c]{0cm}{}	&  spherical galaxy                	&   disk-shaped galaxy    
                                       	&  spherical galaxy                	&   disk-shaped galaxy		\\
      \hline
      \tabhe   	$L$			&  $10^{3}~\mathrm{pc}$   		&   $2.4 \times10^{2}~\mathrm{pc}$	 
					&  $7.0 \times10^{2}~\mathrm{pc}$      	&   $2.4 \times10^{2}~\mathrm{pc}$   	\\
      \tabhe 	$V$	 		&  $3.7~\mathrm{km}~\mathrm{s}^{-1}$    &   $9.7~\mathrm{km}~\mathrm{s}^{-1} $  
					&  $47~\mathrm{km}~\mathrm{s}^{-1}$ 	&   $61~\mathrm{km}~\mathrm{s}^{-1} $ \\
      \tabhe  	$\Rey$	  		&  $1.3 \times10^{11}$     		&   $8.1 \times10^{10}$   
					&  $1.2 \times10^{12}$     		&   $5.1 \times10^{11}$		\\
      \tabhe  	$\Rm$	  		&  $3.0 \times10^{25}$		 	&   $1.9 \times10^{25}$	
					&  $2.7 \times10^{26}$		  	&   $1.2 \times10^{26}$		\\
      \tabhe    $\ell_\nu$ 		&  $1.1 \times10^{-5}~\mathrm{pc}$      &   $3.8 \times10^{-6}~\mathrm{pc}$   
					&  $1.7 \times10^{-6}~\mathrm{pc}$      &   $1.0 \times10^{-6}~\mathrm{pc}$	\\
      \tabhe    $\Gamma$ 		&  $1.5 \times10^{2}~\mathrm{Myr}^{-1}$ &   $1.4 \times10^{3}~\mathrm{Myr}^{-1}$     
					&  $7.1 \times10^{3}~\mathrm{Myr}^{-1}$ &   $1.9 \times10^{4}~\mathrm{Myr}^{-1}$\\
      \tabhe    $t_\nu$	  		&  $1.7 \times10^{-1}~\mathrm{Myr}$  	&   $1.9 \times10^{-2}~\mathrm{Myr}$     
					&  $3.8 \times10^{-3}~\mathrm{Myr}$  	&   $1.5 \times10^{-3}~\mathrm{Myr}$	\\
      \tabhe    $t_L$	  		&  $2.7 \times10^{2}~\mathrm{Myr}$  	&   $24 ~\mathrm{Myr}$	  
					&  $15 ~\mathrm{Myr}$  			&   $3.8 ~\mathrm{Myr}$      \\
      \tabhe    $B_{\mathrm{sat},\nu}$  &  $1.1 \times10^{-9}~\mathrm{G}$  	&   $3.3 \times10^{-9}~\mathrm{G}$    
					&  $7.5 \times10^{-9}~\mathrm{G}$  	&   $1.2 \times10^{-8}~\mathrm{G}$	\\
      \tabhe    $B_{\mathrm{sat},L}$	&  $1.6 \times10^{-6}~\mathrm{G}$  	&   $4.3 \times10^{-6}~\mathrm{G}$	 
					&  $2.1 \times10^{-5}~\mathrm{G}$  	&   $2.7 \times10^{-5}~\mathrm{G}$	\\
    \end{tabular}
\caption{The characteristic quantities of the small-scale dynamo for accretion-driven turbulence (left hand side) and for SN-driven turbulence (right hand side). In each case we present results for a spherical galaxy and a disk-shaped galaxy. We list the forcing scale of the turbulence $L$, the typical turbulent velocity on that scale $V$, the hydrodynamic and magnetic Reynolds numbers $\Rey$ and $\Rm$, the viscous scale $\ell_\nu$, the kinematic growth rate of the dynamo $\Gamma$, the time until saturation on the viscous and the forcing scale occurs $t_\ell$ and $t_L$ and the saturation field strengths on those scales $B_{\mathrm{sat},\nu}$ and $B_{\mathrm{sat},L}$. All the given values in this table are for the fiducial model with a factor $\beta=0.05$ of kinetic energy that goes into turbulent motions and a SN efficiency of $\alpha=0.01$.}
  \label{Table}
\end{table*}

\subsubsection{Accretion-Driven Small-Scale Dynamo}

\begin{figure}[hbtp]
  \centering
  \includegraphics[width=0.48\textwidth]{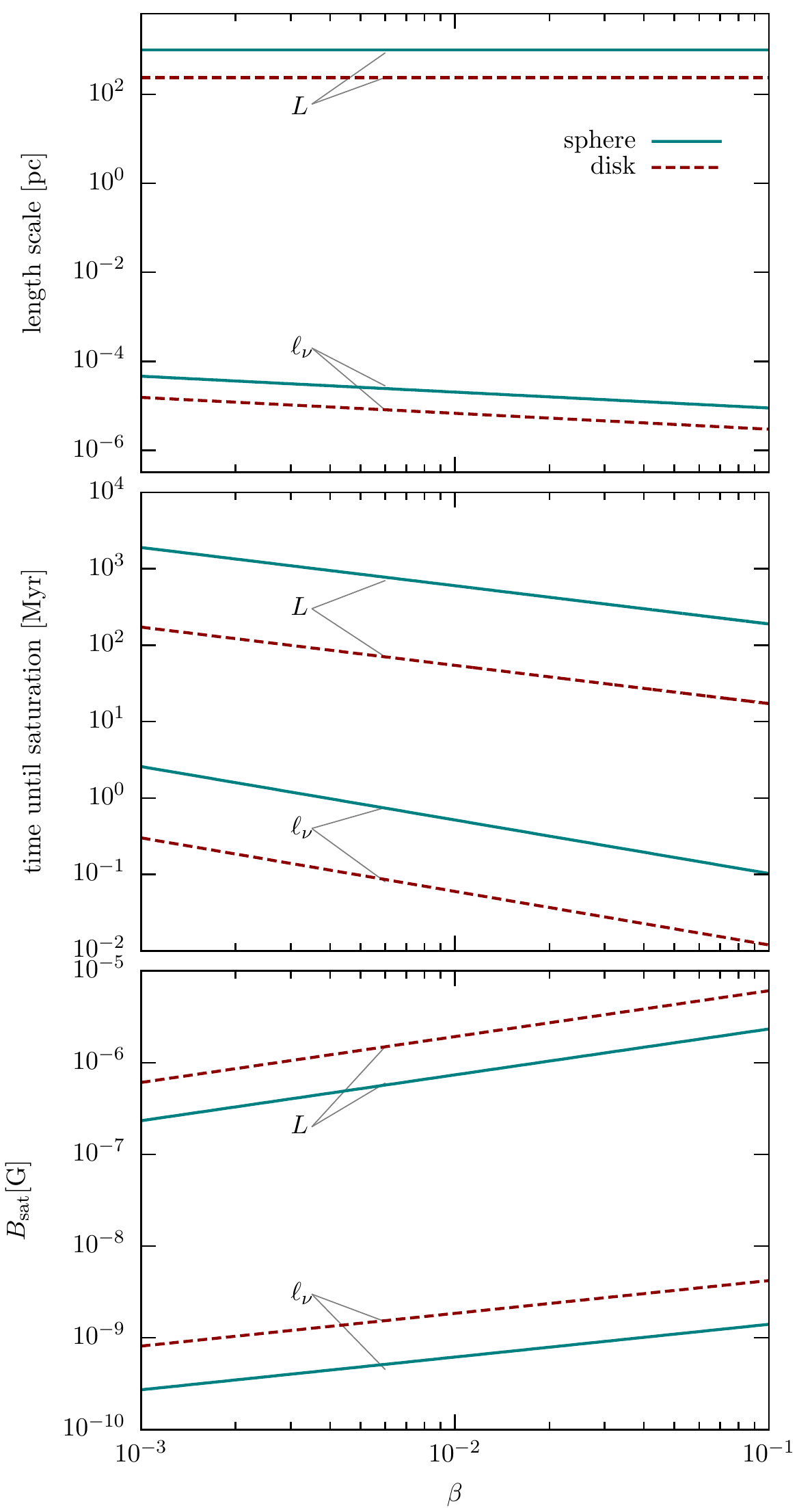}
\caption{The dependency of the accretion-driven small-scale dynamo mechanism on the percentage of kinetic energy that goes into turbulence $\beta$. The upper panel shows the different length scales, the middle panel the time until saturation, i.e.~$t_\nu$ and $t_L$, and the lower panel the saturation magnetic field strength $B_{\mathrm{sat}}$. We plot all quantities on the viscous scale $\ell_\nu$ and on the turbulent forcing scale $L$ as indicated in the figure. The solid blue lines show the results for a spherical galaxy, the dashed red lines the ones for a disk.}
\label{Multiplot_acc}
\end{figure} 

Based on the discussion of the strength of magnetic seed fields in the introduction, we assume the initial magnetic field strength on the viscous scale to be
\begin{eqnarray}
  B_{\nu,0} = 10^{-20}~\mathrm{G}.
\end{eqnarray}
This is a rather conservative estimate. \\
The small-scale dynamo amplifies this seed field as soon as sufficient turbulence has evolved. The typical growth rates in the kinematic phase are summarized in table \ref{Table}. We find 150 Myr$^{-1}$ for the case of a spherical galaxy and 1400 Myr$^{-1}$ for a disk. A fraction of the magnetic energy can be dissipated again by Ohmic diffusion. The dissipation rate on the viscous scale $\ell_\nu$ is given by
\begin{eqnarray}
  \Gamma_{\mathrm{Ohm},\nu} = \frac{\eta}{\ell_\nu^2}.
\end{eqnarray}
In our model $\Gamma_{\mathrm{Ohm},\nu}$ is of the order of $10^{-12}-10^{-10}~\mathrm{Myr}^{-1}$ and thus can be neglected compared the growth rate of the magnetic field.\\
With these relatively large growth rates, the small-scale dynamo amplification works on very short timescales. We find that in a spherical galaxy a magnetic field of $1.6\times10^{-6}$ G and be reached on a scale of $10^3$ pc after 270 Myr. In a disk the saturation field strength is larger by a factor of more than 2. However, the field is only on a scale of 240 pc, but it is saturated after already 24 Myr.\\ 
The efficiency of the small-scale dynamo, i.e.~the saturation magnetic field strength ($B_{\mathrm{sat},\nu}$ or $B_{\mathrm{sat},L}$) that can be achieved and the time on which saturation occurs ($t_\nu$ or $t_L$), depends strongly on the amount of turbulent kinetic energy, controlled by the parameter $\beta$ (see equations \ref{Vacc_sphere} and \ref{Vacc_disk}). In our fiducial model we use $\beta=0.05$, however this is a rough assumption. We test how the dynamo efficiency changes when varying $\beta$ in figure \ref{Multiplot_acc}.\\
In the upper panel of figure \ref{Multiplot_acc} we show the dependency of the viscous scale $\ell_\nu$ and the forcing scale $L$ on $\beta$. Of course $L$ is not effected by $\beta$, while $\ell_\nu$, which is a function of the Reynolds number and thus of the turbulent velocity, decreases with increasing $\beta$. The time until saturation of the dynamo, which is shown in the middle panel of figure \ref{Multiplot_acc}, also decreases with increasing $\beta$. This is a natural consequence of the larger amount of turbulent kinetic energy. In the same way the plot in the lower panel can be understood: the more turbulent energy, i.e.~the higher $\beta$, the higher is the saturation field strength. The magnetic field strength on the forcing scale $B_{\mathrm{sat},L}$ increases as
\begin{eqnarray}
  B_{\mathrm{sat},L} \propto \beta^{1/2}.
\end{eqnarray}

\vfill
\eject

\subsection{Magnetic Fields from Stellar Feedback}

\subsubsection{Distributing Stellar Magnetic Fields by Supernovae}

A natural source for magnetic fields in the ISM of galaxies are stellar magnetic fields that get distributed over large volumes by SN explosions. \citet{SchoberEtAl2012.2} have shown that the small-scale dynamo can produce strong magnetic fields during primordial star formation. Hints to dynamical important magnetic fields during the formation of the first stars come also from high-resolution numerical simulations \citep{FederrathEtAl2011.1,TurkEtAl2012,SurEtAl2012} and further semi-analytical calculations \citep{SchleicherEtAl2010}. Thus, we expect the first and second generations of stars to be magnetized.

\paragraph{Properties of Supernova Candidates}
We assume that a typical star that ends in a supernova has a mass of
\begin{eqnarray}
  M_\mathrm{star} = 10~M_\odot
\end{eqnarray} 
and a radius of 
\begin{eqnarray}
  R_\mathrm{star} = \left(\frac{M_\mathrm{star}}{M_\odot}\right)^{0.8} R_\odot
\end{eqnarray}  
with the solar mass $M_\odot = 2 \times 10^{33}~\mathrm{g}$ and radius $R_\odot = 7 \times 10^{5}~\mathrm{km}$.\\
It is very difficult to estimate the magnetic energy in a typical population III star, as there is not much theoretical work on that topic so far. In principle, one could assume that a certain percentage of the total energy of the SN energy is within the magnetic field. If the magnetic energy $B_\mathrm{star}^2/(8 \pi)~4/3 \pi R_\mathrm{star}^3$ equals e.g.~0.001 $E_\mathrm{SN}$, the stellar magnetic field would have a very high value of $B_\mathrm{star} = 8\times10^6~\mathrm{G}$.\\
Here, however, we use as a crude estimate for the magnetic field of population III stars based on observations of present-day massive stars. In most high-mass stars no magnetic fields are detected, there are few percent of stars with an enhanced magnetic field (see \citet{DonatiLandstreet2009}). These so-called ``peculiar A or B'' stars have a typical dipole field strength of
\begin{eqnarray}
  B_\mathrm{star} = 10^4~\mathrm{G}.
\end{eqnarray}
We take this value as an upper limit of magnetic fields in primordial stars, but test also lower stellar field strengths in the following.

\paragraph{Evolution of a Supernova Remnant}
\begin{figure}[hbtp]
  \centering
  \includegraphics[width=0.48\textwidth]{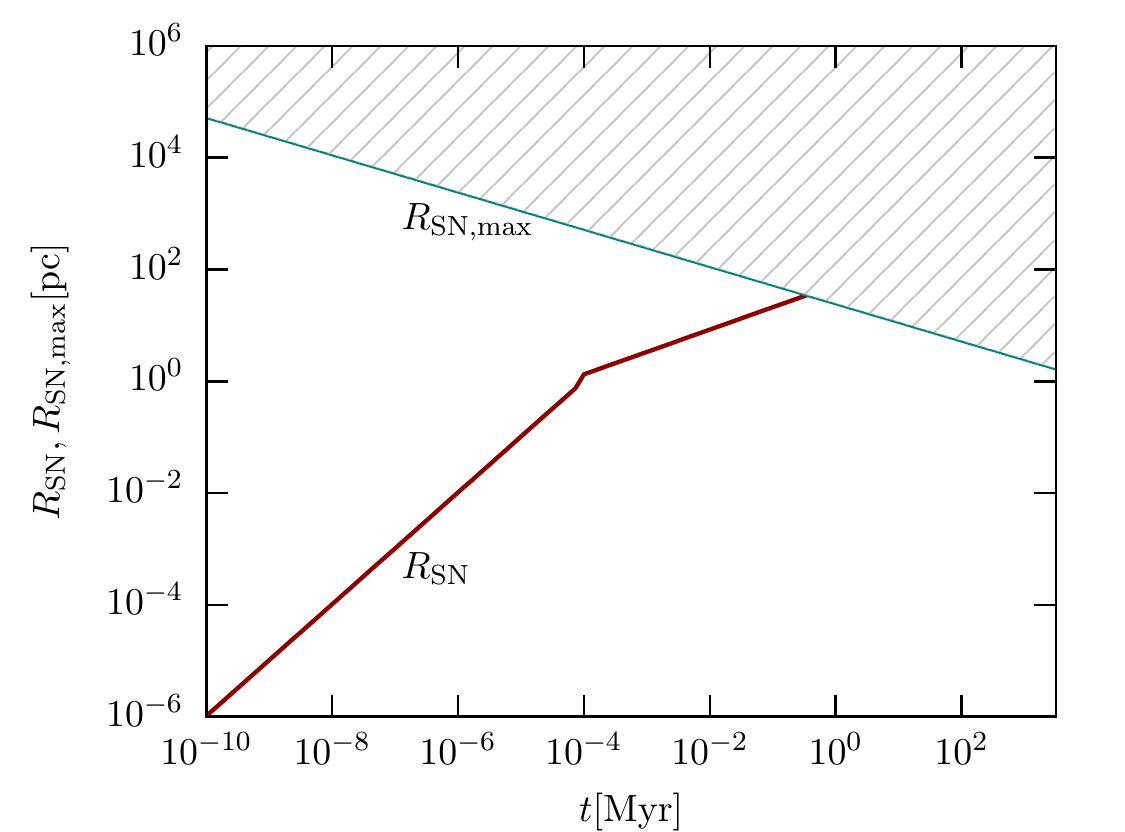}
\caption{The red line shows the radius of a SN shock $R_\mathrm{SN}(t)$ as a function of time. Up to roughly $100$ yr the SN shock expands freely, then the Sedov-Taylor expansion sets in. The available maximum radius for SNe $R_\mathrm{SN,max}(t)$ as a function of time is shown for the case of a spherical halo by the blue line. This radius decreases in time, as the number of SN in the protogalactic core increases. The first SN shocks collide after a time of approximately $0.36~\mathrm{Myr}$.}
\label{RSNR_t}
\end{figure}
Stars with masses above 8 solar are expected to explode as a core-collapse supernova, introducing additional turbulent energy into the ISM \citep{Choudhuri1998,Padmanabhan2001}. Initially the shock front of a SN expands freely, i.e.~the pressure of the surrounding ISM is negligible. The shock velocity $v_\mathrm{e}$ can then be determined by
\begin{eqnarray}
  E_\mathrm{SN} = \frac{1}{2} M_\mathrm{e} v_\mathrm{e}^2,
\end{eqnarray}
where $E_\mathrm{SN}$ is the energy of a SN (neglecting the energy loss by neutrinos) and $M_\mathrm{e}$ the ejected mass. The shock radius $R_\mathrm{SN}$ as a function of time $t$ is thus
\begin{eqnarray}
  R_\mathrm{SN}(t) = \left(\frac{2 E_\mathrm{SN}}{M_\mathrm{e}}\right)^{1/2}~t.
\end{eqnarray}
The free expansion phase ends, when the accumulated mass of the ISM in front of the shock is of order of $M_\mathrm{e}$. This happens at the so-called sweep-up radius $R_\mathrm{sw}$ defined by 
\begin{eqnarray}
  M_\mathrm{e} = \frac{4}{3} \pi R_\mathrm{sw}^3 \rho
\end{eqnarray}
with $\rho$ being the mean density of the ISM. The shock front reaches $R_\mathrm{sw}$ at a time
\begin{eqnarray}
  t_\mathrm{sw} = \frac{R_\mathrm{sw}}{v_\mathrm{e}} = \left(\frac{3}{4 \pi \rho}\right)^{1/3} \left(\frac{1}{2 E_\mathrm{SN}}\right)^{1/2} M_\mathrm{e}^{5/6},
\end{eqnarray}
which is in our model of the order of 100 yr. For $t > t_\mathrm{sw}$ the expansion of the supernova remnant is driven adiabatically by thermal pressure, which is known as the Sedov-Taylor phase \citep{Sedov1946,Taylor1950,Sedov1959}. We can estimate the radius of the shock in this case with
\begin{eqnarray}
  \frac{\mathrm{d}}{\mathrm{d}t} \left(4 \pi R_\mathrm{SN}^3 \rho \dot{R}_\mathrm{SN}\right) = 4 \pi R_\mathrm{SN}^2 P,
\label{SedovTaylor}
\end{eqnarray}
where the $\dot{~}$ indicates a time derivative and the pressure $P$ is given by
\begin{eqnarray}
  P = (\gamma -1) \frac{E_\mathrm{SN}}{\frac{4}{3} \pi R_\mathrm{SNR}^3}
\end{eqnarray}
with $\gamma = 5/3$ for adiabatic expansion. We can solve equation (\ref{SedovTaylor}) with a simple power-law ansatz and find
\begin{eqnarray}
  R_\mathrm{SN}(t) = \left(\frac{25 E_\mathrm{SN}}{4 \pi \rho}\right)^{1/5}~t^{2/5}.
\end{eqnarray}
Thus, the evolution of the SN remnant can be described by \citep{Choudhuri1998}
\begin{eqnarray}
  R_\mathrm{SN}(t) = \begin{cases} \left(\frac{2 E_\mathrm{SN}}{M_\mathrm{e}}\right)^{1/2}~t        & t < t_\mathrm{sw} \\ 
  			           \left(\frac{25 E_\mathrm{SN}}{4 \pi \rho}\right)^{1/5}~t^{2/5}   & t > t_\mathrm{sw}. 
  	             \end{cases}
\label{RSN}
\end{eqnarray}
We assume now that the energy released in a SN explosion is $E_\mathrm{SN} = 10^{51}~\mathrm{erg}$ and that about 10 percent of the mass of the progenitor star is ejected, i.e.~$M_\mathrm{e} \approx 0.1~M_\mathrm{star}$. In our model the SN remnants evolve as described in equation (\ref{RSN}) and shown in figure \ref{RSNR_t} until they collide. At later stages of shock evolution other energy loss mechanisms become dominant. The electrons lose their energy by ionization, bremsstrahlung, synchrotron emission and inverse Compton scattering. The latter is the most important energy loss channel at high redshifts as here the density of the CMB photons is considerably larger \citep{SchleicherBeck2013}.\\
If the SNe are distributed homogeneously in the protogalaxy, each SN shell has a mean maximum radius at the first collision of
\begin{eqnarray}
  R_\mathrm{SN,max}(t) = \frac{R}{N_\mathrm{SN}(t)^{\xi}},
\label{RSNmax}
\end{eqnarray}
where the radius of the galaxy $R$ is given in equations (\ref{Rsph}) and (\ref{Rdisk}) and the exponent $\xi$ depends on the geometry of the galaxy. In case of a spherical halo $\xi=1/3$, in case of a thin disk $\xi=1/2$. The maximum expansion radius of the SN shock is shown in figure \ref{RSNR_t} for the spherical case.\\ 
By comparing (\ref{RSN}) to (\ref{RSNmax}) we find the typical time scale for SN collisions $t_\mathrm{SN}$. At that point the SN bubbles fill approximately the whole galaxy. In the spherical case we find
\begin{eqnarray}
  t_\mathrm{SN} \approx 0.36~\mathrm{Myr},
\end{eqnarray}
which we take as the typical timescale for SN collisions. Further, we use $R_\mathrm{SN}(t_\mathrm{SN})$ as the typical length scale of SN shocks.

\begin{figure}[hbtp]
  \centering
  \includegraphics[width=0.48\textwidth]{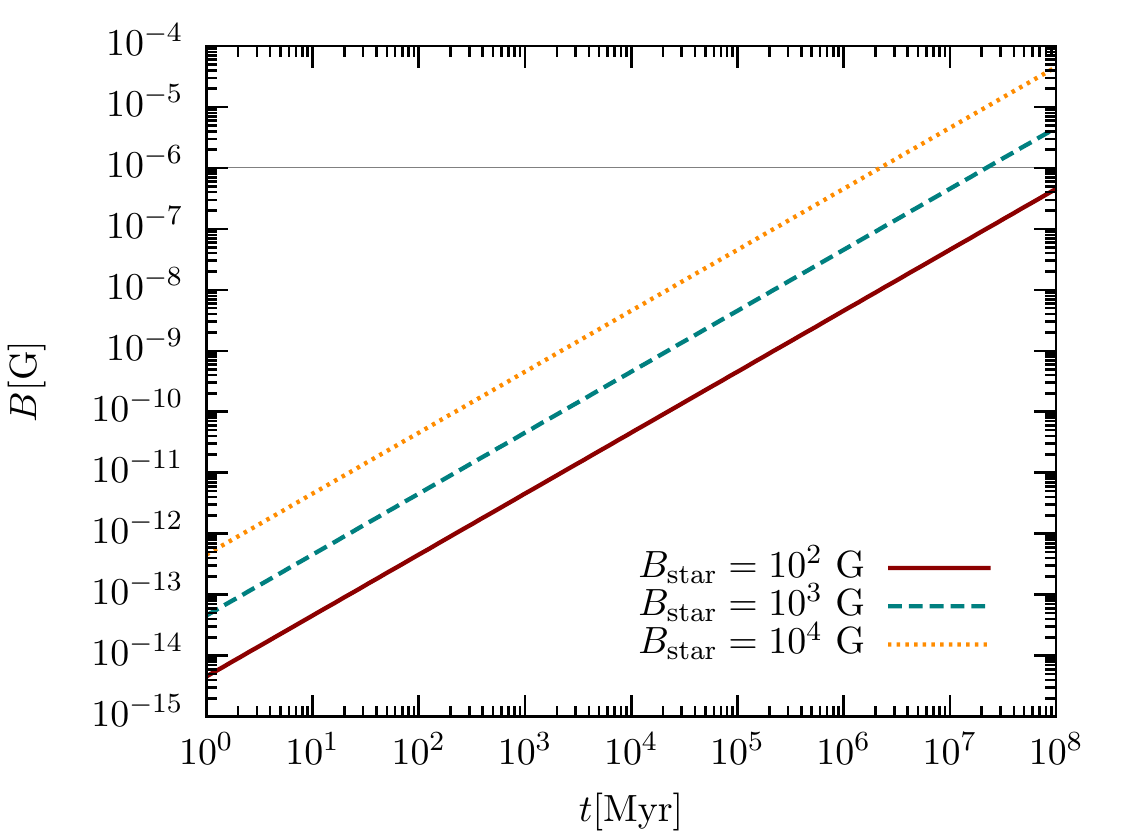}
\caption{The evolution of the magnetic field, when the only source of magnetic energy in the ISM are stellar magnetic fields distributed by SN explosions. The curves show the results for a spherical halo in our fiducial model with a SN efficiency of $\alpha=0.01$. We show three different mean stellar field strengths: $10^{4}$ G (yellow dotted line), $10^{3}$ G (green dashed line) and $10^{2}$ G (solid red line). With the thin gray line we further indicate the typical saturation strength of a magnetic field generated by a small-scale dynamo.}
\label{Bfield_SN}
\end{figure}

\paragraph{Magnetic Field Evolution}
If now all the stellar magnetic energy is distributed into the volume available by the SN explosion and no significant magnetic energy is left in the stellar remnant, the resulting magnetic field strength in the ISM after the first SN generation is
\begin{eqnarray}
  B_\mathrm{ISM} = \left(\frac{R_\mathrm{star}}{R_\mathrm{SN}(t_\mathrm{SN})}\right)^{2}~B_\mathrm{star}.
\end{eqnarray}
Here, we assumed a spherical shape of the galaxy and flux freezing. All the following stars will produce roughly the same amount of magnetic energy, that is then distributed in the ISM by SN. Thus, the time evolution of the stellar magnetic fields in galaxies can be approximated by
\begin{eqnarray}
  B_\mathrm{ISM}(t) = \left(\frac{R_\mathrm{star}}{R_\mathrm{SN}(t_\mathrm{SN})}\right)^{2}~B_\mathrm{star}~\frac{t}{t_\mathrm{SN}}.
\end{eqnarray}
The values of $R_\mathrm{SN}(t_\mathrm{SN})$ and $t_\mathrm{SN}$ depend obviously on the SN rate, which is determined by the parameter $\alpha$ as defined in (\ref{SNR}). We obtain for the spherical case:
\begin{eqnarray}
  t_\mathrm{SN} \propto \alpha^{-5/11} \\
  R_\mathrm{SN}(t_\mathrm{SN}) \propto \alpha^{-2/11}
\end{eqnarray}
leading to a dependency of the magnetic field distributed by SN on the efficiency of the SN rate of
\begin{eqnarray}
  B_\mathrm{ISM}(t) \propto \alpha^{9/11}~t.
\end{eqnarray}
In figure \ref{Bfield_SN} we show the evolution of the distributed magnetic fields for different mean magnetic fields of the stars ($10^4$ to $10^2$ G) and for our fiducial case of $\alpha \approx 0.01$. Note that the case of $10^4$ G is an upper limit of magnetic fields in massive stars. We assume the magnetic fields of the first stars to be considerably lower. \\
The distribution of stellar magnetic fields by SNe explosions thus seems to be not important compared to the dynamo amplification in the ISM. However after a sufficient time the SN could contribute to the magnetic energy in the ISM. If the equipartition field strength is roughly $10^{-6}$ G, the time after which SN distribution becomes important is
\begin{eqnarray}
 t \approx t_\mathrm{SN}~\left(\frac{R_\mathrm{SN}(t_\mathrm{SN})}{R_\mathrm{star}}\right)^{2}~\left(\frac{10^{-6}~\mathrm{G}}{B_\mathrm{star}}\right).
\end{eqnarray}
For our fiducial model we find that this time is about $2.2 \times10^6$ Myr in case of typical stellar field strengths of $10^4$ G. Observations of present-day massive stars indicate that only a few percent have these high field strengths. We thus also consider the more likely case of lower mean stellar fields. For a mean strength of $10^3$ G we find that a micro-Gauss ISM field is only reached after $2.2 \times10^7$ Myr and for a mean strength of $10^2$ G after $2.2 \times10^8$ Myr. Thus, the typical timescales of distribution of stellar magnetic fields by supernovae exceed the age of the Universe by many orders of magnitude and this process cannot be an important contribution for the fields in the ISM, unless the first stars were much stronger magnetized than the present-day stars.\\
In the case of a flat disk-shaped galaxy, where we assume the parameter $\xi$ in equation (\ref{RSNmax}) to be 1/2, the distribution of stellar magnetic fields proceeds marginally faster. Here the typical time until a field strength of $10^{-6}$ G in the ISM is reached is roughly a factor of 10 shorter. \\
In reality the evolution of magnetic fields in SN shock fronts is of course more complicated. In addition to simple flux freezing further amplification processes can take place. \citet{MirandaOpherOpher1998} argue that in a multiple explosion scenario of structure formation \citep{OstrikerCowie1981,MirandaOpher1997} magnetic seed fields of the order of $10^{-10}$ G can be produced on galactic scales. In their model a Biermann battery is operating in the shock of SN explosions of the first stars as here non-parallel gradients of temperature and density can be established. Recently also \citet{BeckEtAl2013} have analyzed the magnetic field evolution in protogalaxies based on SN explosions with the cosmological N-body code GADGET. They find that a combination of SNe and subsequent magnetic field amplification leads to magnetic field strengths of the order of a few $\mu$G, which is comparable to our results, and that the strength of seed field is coupled to the star formation process.

\subsubsection{Dynamo Amplification Driven by SN Turbulence}

\begin{figure}[h!]
  \centering
  \includegraphics[width=0.48\textwidth]{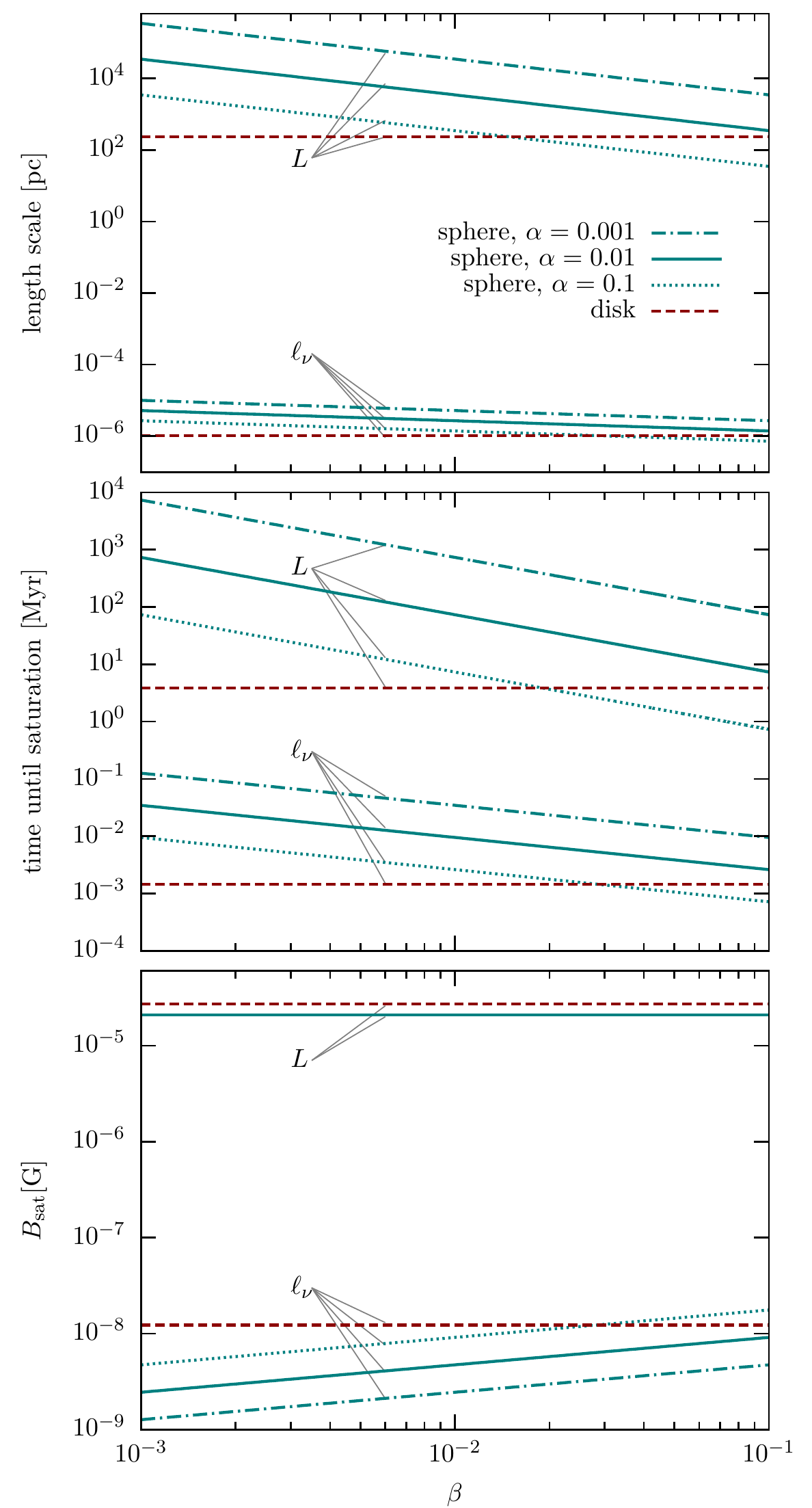}
\caption{The dependency of the SN-driven small-scale dynamo mechanism on the percentage of kinetic energy that goes into turbulence $\beta$ and the SN efficiency $\alpha$. The upper panel shows the different length scales, the middle panel the time until saturation, i.e.~$t_\nu$ and $t_L$, and the lower panel the saturation magnetic field strength $B_{\mathrm{sat}}$. We plot all quantities on the viscous scale $\ell_\nu$ and on the turbulent forcing scale $L$ as indicated in the figure. The dashed-dotted blue line represents the case of a spherical galaxy with $\alpha = 0.001$, the solid blue line the fiducial case of $\alpha = 0.01$ and the dotted blue line the case of $\alpha = 0.1$. The dashed red line shows the results for a disk-like galaxy. There are only 6 lines in the lower plot instead of 8. This results from the fact, that in the spherical case the saturation field strength on the forcing scale does not depend on $\alpha$ nor on $\beta$, in contrary to $L$ and $t_L$ (see equations (80) to (\ref{quantities})).}
\label{Multiplot_SN}
\end{figure} 

\paragraph{SN-Driven Dynamo in a Spherical Galaxy}

In section \ref{SubSec_Turbulence} we discussed the generation of SN turbulence based on numerical simulations. Now we estimate the typical forcing scale $L_\mathrm{SN}$ and the fluctuation velocity on that scale $V_\mathrm{SN}$ in order to determine the Reynolds number (\ref{Re}) and the resulting growth rate of the kinematic small-scale dynamo (\ref{Gamma}). For that we assume that the turbulence driving in the galaxy is in equilibrium.\\
Then the turbulent pressure, which is roughly
\begin{eqnarray}
  P_\mathrm{turb} \approx V_\mathrm{SN}^2 \rho,
\label{Pturb}
\end{eqnarray}
balances the hydrostatic pressure $P$ determined by 
\begin{eqnarray}
  \frac{\mathrm{d}P}{\mathrm{d}r} = \rho~g(r).
\label{P}
\end{eqnarray}
The gravitational acceleration in the spherical case is roughly $g(r) = \frac{4}{3} \pi \rho G r$. Solving equation (\ref{P}) and setting it equal to equation (\ref{Pturb}) yields the turbulent velocity $V_\mathrm{SN}$. We find in the spherical case:
\begin{eqnarray}
  V_\mathrm{SN} \approx \left(\frac{2}{3} \pi \rho G \right)^{1/2} R_\mathrm{sph}.
\end{eqnarray}
The forcing length scale can be estimated by comparing the energy input rate with the dissipation rate:
\begin{eqnarray}
  SNR~\beta~E_\mathrm{SN} = \frac{\frac{1}{2} \rho V_\mathrm{SN}^2}{t_\mathrm{dis}},
\end{eqnarray}
where the dissipation timescale is
\begin{eqnarray}
  t_\mathrm{dis} = \frac{L_\mathrm{SN}}{V_\mathrm{SN}}
\end{eqnarray}
and $SNR$ is the supernova rate (\ref{SNR}). Thus, we find the typical forcing scale of SN-driven turbulence
\begin{eqnarray}
  L_\mathrm{SN} = \frac{\rho~V_\mathrm{SN}^3}{SNR~\beta~E_\mathrm{SN}}.
\label{LSN}
\end{eqnarray}
As in the case of the accretion-driven small-scale dynamo we start with an initial magnetic field strength on the viscous scale of 
\begin{eqnarray}
  B_{\nu,0} = 10^{-20}~\mathrm{G}.
\end{eqnarray}
The turbulence driven by SN shocks makes dynamo action possible, which leads to rapid amplification of the seed field according to equations (\ref{Bnu}) and (\ref{BL}). For the case of a spherical galaxy we find that the growth rate in the kinematic amplification phase is $7.1\times10^{3}~\mathrm{Myr}^{-1}$. After a time of $15~\mathrm{Myr}$ the saturation field strength of $2.1\times10^{-5}~\mathrm{G}$ is reached on the forcing scale. The characteristic quantities of our fiducial models for the SN-driven dynamo are summarized in the right part of table \ref{Table}. \\
As in case of accretion turbulence, the efficiency of the small-scale dynamo is sensible to the amount of kinetic energy that goes into turbulence $\beta$. Moreover, when modeling the scale of turbulence forcing we add another uncertainty namely the $SNR$, which includes the efficiency parameter $\alpha$ (see equation \ref{SNR}). In our fiducial model we choose $\alpha=0.01$, but there could easily be a variation of a factor 10. The dependency of the quantities most important for the dynamo amplification on $\alpha$ and $\beta$ in case of a spherical halo is the following:
\begin{eqnarray}
  \ell_{\nu,\mathrm{SN}} & \propto & \left(\alpha \beta\right)^{-0.28} \\
  L_\mathrm{SN} & \propto & \left(\alpha \beta\right)^{-1} \\
  V_\mathrm{SN} & = & \mathrm{const} \\
  B_{\nu,\mathrm{sat,SN}} & \propto & \left(\alpha \beta\right)^{0.28} \\
  B_{L,\mathrm{sat,SN}} & = & \mathrm{const} 
\label{quantities}
\end{eqnarray} 
We show the dependency of the length scales, the time until saturation and the saturation magnetic field strength on $\beta$ and for different values of $\alpha$ in figure \ref{Multiplot_SN}.

\paragraph{SN-Driven Dynamo in a Disk-Like Galaxy}
We perform the same analysis for the disk case. Here the gravitational acceleration becomes independent of the radius for a thin disk, i.e.~$H_\mathrm{disk} \ll r$. In that approximation we find $g(r) \approx 2 \pi \rho G H_\mathrm{disk}$, which leads to a turbulent velocity of
\begin{eqnarray}
  V_\mathrm{SN} \approx \left(2 \pi \rho G R_\mathrm{disk} H_\mathrm{disk}\right)^{1/2}.
\end{eqnarray}
The forcing scale can be determined by equation (\ref{LSN}). In case of a disk-shaped galaxy we find that the typical forcing scale $L_\mathrm{SN}$ 
\begin{eqnarray}
  L_\mathrm{SN} \approx H_\mathrm{disk}.
\end{eqnarray}
We find that the kinematic growth rate in our fiducial model is $1.9\times10^{4}~\mathrm{Myr}^{-1}$. The time until saturation on the forcing scale is then only 3.8 Myr and the saturation field strength is $2.7\times10^{-5}~\mathrm{G}$. \\
In case of a disk-shaped galaxy all the quantities (79) to (\ref{quantities}) are independent of $\alpha$ and $\beta$. For comparison with the spherical galaxy we show them however also in figure \ref{Multiplot_SN}. 

\vfill
\eject

\section{Conclusions}
\label{Sec_Conclusions}

In this paper we model the evolution of the (turbulent) magnetic field in a young galaxy. We find that weak magnetic seed fields get amplified very efficiently by the small-scale dynamo (see table \ref{Table}), which is driven by turbulence from accretion and from supernova (SN) explosions. Dynamo theory predicts that the magnetic field is amplified in two phases: in the kinematic phase the field grows exponentially until the dynamo is saturated on the viscous scale. Then the non-linear phase begins, where the magnetic energy is shifted towards larger scales until saturation on the turbulent forcing scale occurs. \\
For our fiducial models of a young galaxy we use a fixed particle density of $10~\mathrm{cm}^{-3}$ and a temperature of $5 \times 10^3~\mathrm{K}$. We concentrate on two different geometries: a spherical and a disk-shaped galaxy (see section \ref{Sec_Model_General}). We determine the viscosity of the plasma, which becomes anisotropic when the plasma becomes magnetized, and the magnetic diffusivity. Turbulence is generated by accretion flows onto the galactic core and also by SN shocks. By estimating typical driving scales and velocities we can determine the hydrodynamic and the magnetic Reynolds number. The magnetic field evolution depends strongly on the type of turbulence, which we assume to be a mixture of solenoidal and compressive modes. \\
For our fiducial model we find that the dynamo saturates on the largest scale in accretion-driven turbulence after a time of roughly $270~\mathrm{Myr}$ in case of a spherical galaxy and after $24~\mathrm{Myr}$ in case of a disk. Turbulence generated by SN shocks can amplify the magnetic field on shorter timescales, with saturation reached after $15~\mathrm{Myr}$ in a spherical galaxy and $3.8~\mathrm{Myr}$ in a disk. The dynamo timescale is thus comparable to the free-fall time $t_\mathrm{ff}=\left(3 \pi/(32 G \rho)\right)^{1/2} \approx 16~\mathrm{Myr}$. The age of the Universe at the onset of galaxy formation, i.e.~at a redshift of 10, is roughly 470 Myr, which is larger than the dynamo timescales by factor of 2 to 120 for our four fiducial models. In the models with the longest amplification times our assumption of constant accretion and supernova rates may thus not be very precise. Nevertheless, these models provide an order of magnitude estimate of the resulting magnetic strength. In case of a disk-like galaxy we can compare the dynamo timescales further to the typical time of one rotation, which turns out to be $340~\mathrm{Myr}$ when using the Kepler velocity (\ref{VKepler}). Thus, we can expect that the small-scale dynamo is saturated within less then one orbital time, the turbulent magnetic field gets ordered and an $\alpha-\Omega$ dynamo, i.e.~a galactic large-scale dynamo, sets in. The role of rotation has been analyzed for example by \cite{KotarbaEtAl2009} and \cite{KotarbaEtAl2011} in numerical simulations and is extremely important for understanding the present-day large-scale structure of galactic magnetic fields. \\
The magnetic field strengths predicted by our fiducial models are very high with values between $1.6\times10^{-6}~\mathrm{G}$ and $4.3\times10^{-6}~\mathrm{G}$ in the accretion-driven case and between $2.1\times10^{-5}~\mathrm{G}$ and $2.7\times10^{-5}~\mathrm{G}$ in the SN-driven case for a spherical galaxy and a disk, respectively. These field strengths are comparable with the ones observed in the local Universe, where the typical turbulent field component in present-day disk galaxies is $(2-3)\times10^{-5}~\mathrm{G}$ in spiral arms and bars and up to $(5-10)\times10^{-5}~\mathrm{G}$ in the central starburst regions \citep{Beck2011}. New radio observations detect also magnetic fields in dwarf galaxies. Their field strengths, which seem to be correlated with the SFR, are typically a factor of roughly three lower compared to the one in spiral galaxies \citep{ChyzyEtAl2011}. \\
Our calculations suggest that the turbulent magnetic field of a galaxy was very high already at high redshifts. An observational confirmation of this result is very complicated. A hint towards an early generation of the turbulent magnetic field in galaxies comes from \citet{HammondRobishawGaensler2012}. They analyze the rotation measure of a huge catalog of extragalactic radio sources as a function of redshift and find that it is constant up to redshifts of 5.3, which is the maximum redshift in their dataset. A very powerful tool provides moreover the far-infrared (FIR) - radio correlation, which relates the star formation rate to the synchrotron loss of cosmic ray electrons. It is observed to be constant up to redshifts of roughly 2 \citep{SargentEtAl2010, BourneEtAl2011}, but is expected to break down at a higher redshift, which depends on the star formation rate and the evolution of typical ISM densities \citep{SchleicherBeck2013}. With new instruments like SKA and LOFAR our knowledge about the evolution of the cosmic magnetic fields will increase.  \\
Besides our fiducial models we analyze the effect of changing the amount of kinetic energy that goes into turbulence and find that the dynamo is more efficient with increasing turbulent energy, which is intuitively clear. Furthermore we determine the small-scale dynamo evolution for a varying supernova rate ($SNR$), which is important for estimating the driving scale of SN turbulence in the case of a spherical core. As expected the time until saturation increases with increasing $SNR$. However, the typical largest scale of the magnetic field decreases with the $SNR$. \\  
We further estimate the effect of magnetic field enrichment in galaxies by distributing stellar fields by SN explosions. As an estimate of the magnetic energy in the first stars is very hard, we determine the expected magnetic field evolution in the ISM for three different cases. An upper limit of magnetic field strengths of the primordial stars is $10^4$ G, which is a value observed in a the few percent of present-day massive stars that are magnetized. Distributing these mean stellar fields by SNe in the ISM, we find that a ISM field strength of $10^{-6}$ G is reached after $10^6$ Myr, which is already longer than the Hubble time. Thus, the dynamo increases the magnetic field strength much faster. \\
With our model we have shown that the small-scale dynamo can amplify weak magnetic seed fields in the ISM of early galaxies on relatively short time scales compared to other evolutionary timescales. This leads to the build-up of strong magnetic fields already at very early phases of (proto)galactic evolution. Theoretical models of galaxy evolution describe a collapse of a spherical object to a disk. Comparison of the gravitational energy, $3/5 G M^2 R^{-3} \approx 10^{55}~\mathrm{erg}$, with the magnetic energy at dynamo saturation, $B^2/(8 \pi) \approx 10^{53}~\mathrm{erg}$, shows that the field is not strong enough to prevent to collapse. Still the magnetic field produced by the small-scale dynamo has potentially strong impact on ISM dynamics and subsequent star formation.

\acknowledgements
{We thank the anonymous referee for useful comments on our manuscript. We acknowledge funding through the {\em Deutsche Forschungsgemeinschaft} (DFG) in the {\em Schwer\-punkt\-programm} SPP 1573 ``Physics of the Interstellar Medium'' under grant KL 1358/14-1 and SCHL 1964/1-1. Moreover, we thank for financial support by the {\em Baden-W\"urttemberg-Stiftung} via contract research (grant P-LS-SPII/18) in their program ``Internationale Spitzenforschung II'' as well as the DFG via the SFB 881 ``The Milky Way System'' in the sub-projects B1 and B2. J.~S. acknowledges the support by IMPRS HD. D.~R.~G.~S.~thanks for funding via the SFB 963/1 (project A12) on ``Astrophysical flow instabilities and turbulence''.}
~\\
~\\
~\\

\vfill
\eject

\end{document}